\documentclass[12pt,a4paper,aps]{revtex4}
\usepackage{epsfig}
\begin{document}
\title{Inhomogeneous DNA: conducting exons and insulating introns}
\author{A.A.~Krokhin$^{1}\footnote{Corresponding author: arkady@unt.edu}$, V. M. K.~Bagci$^{1}$,
F.M.~Izrailev$^{2}$, O.V.~Usatenko$^3$, and V.A.~Yampol'skii$^3$}
\address{$^1$Department of Physics, University of North Texas, P.O.
Box 311427, Denton, TX
76203}
\address{$^2$Instituto de F{\'\i}sica, Universidad Aut\'onoma de Puebla, Apartado
Postal J-48,
Puebla, 72570 Mexico}
\address{$^3$A. Ya. Usikov Institute for Radiophysics and
Electronics, Ukrainian Academy of Science, 12 Proskura Street, 61085 Kharkov, Ukraine}
\begin{abstract}
Parts of DNA sequences known as exons and introns play very
different role in coding and storage of genetic information. Here
we show that their conducting properties are also very different.
Taking into account long-range correlations among four basic
nucleotides that form double-stranded DNA sequence, we calculate
electron localization length for exon and intron regions.
Analyzing different DNA molecules, we obtain that the exons have
narrow bands of extended states, unlike the introns where all the
states are well localized. The band of extended states is due to a
specific form of the binary correlation function of the sequence
of basic DNA nucleotides.

\end{abstract}
\date{\today}

\maketitle

\section{Introduction} A DNA molecule is an exciting example of a
natural complex system with intriguing properties. Many of these
properties remain unexplained and need new approaches for  further
analysis. One of the fundamental questions is how information is
transferred along a sequence of nucleotides. For example, if a
mutation occurs in the sequence, it is usually healed. This means
that some of physical parameters of the DNA molecule are
sufficiently sensitive to detect this mutation. The length of a
mutation is relatively short ($\sim 10$ base pairs) as compared to
the length of a gene ($\sim 10^3-10^6$ base pairs). Because of
small statistical weight of a mutation, the mechanical and
thermodynamic characteristics are not sensitive enough for its
robust detection. Unlike this, the electrical resistance of a DNA
molecule strongly fluctuates even if a single nucleotide in a long
sequence is replaced (or removed) \cite{muta,Guo}. This property
is a signature of coherent transport that gives rise to universal
fluctuations of conductance in mesoscopics samples \cite{alt}.

In a DNA molecule the charge carriers move along a double-helix
formed by two complementary sequences of four basic nucleotides:
A, T, G, and C. A conduction band would form, if the DNA texts
would exhibit some periodicity \cite{Bri}. However, many studies
of the DNA texts have revealed rich statistical properties but not
the periodicity. One of the suggestions is that a DNA molecule is
a stochastic sequence of nucleotides, the main feature of which is
long-range correlations \cite{corr}. Therefore a popular method of
detection of correlations is mapping of a DNA sequence onto a
random walk. Long-range correlations are manifested then in an
anomalous scaling of the generated classical diffusion
\cite{Peng}.

Quantum transport through a DNA molecule is also strongly affected
by the correlations. An uncorrelated sequence of nucleotides
localizes all quantum electron states, as occurs in any 1D
white-noise potential, making impossible charge transfer at
distances longer than the localization length $l(E)$. However,
since most of the mutations in DNA are successfully healed,  one
may assume the existence of charge transport \cite{bio} through
{\it delocalized} states that are responsible for the transfer of
information at much longer distances. Such delocalized states are
expected to exist within exons -- the coding regions where the
genetic information is stored. An important feature of charge
transfer in carrying mutations exons was reported in Ref.
\cite{Rom2}. It was shown that cancerous mutations usually produce
much less variation in the resistance than noncancerous ones. This
apparent distinction shed light on the problem of survival of
cancerous mutations. The healing of a mutation occurs only if it
is detected by base excision repair enzymes. Since the detection
of the mutation is most likely due to DNA-mediated charge
transport \cite{bio1}, it is clear that cancerous mutations, being
"electrically masked," are very unlikely to be detected and then
repaired.

On the other hand, the introns -- the long segments of DNA that
apparently do not carry genetic code -- may not contain
delocalized states in the energy spectrum, thus remaining
insulators. In this Letter we give evidence for the validity of
this hypothesis using a theoretical approach based on the results
of electron localization in correlated disordered potentials. Our
study  of various DNA molecules shows that the energy spectrum of
the exons indeed contains practically delocalized states. Unlike
this, the electron wave functions are well-localized within the
introns.

\section{Two-stranded model of DNA} Let us first consider the
widely used model for electron transport in DNA molecules,  that
is a discrete lattice with random on-site potential $\epsilon_n$
and site-independent nearest-neighbor hopping amplitude $t$,
\begin{equation}
\label{se1}t(\psi _{n+1}+\psi _{n-1})=(\epsilon _n-E)\,\psi _n\,.
\end{equation}
The energies $\epsilon_n$ are the ionization energies of the four
nucleotides, $\epsilon_A=8.24$, $\epsilon_T=9.14$,
$\epsilon_C=8.87$, and $\epsilon_G=7.75$ eV,  and the hopping
amplitude $t$ may vary from 0.1 to 1 eV \cite{Enders}. Although
the on-site energies in a sequence of {\it coupled} nucleotides do
not coincide exactly with their ionization potentials,  one may
neglect this difference as it plays a minor role in our
consideration.  The regular periodic potential $\epsilon_n = V_0$
in Eq. (\ref{se1}) gives rise to Bloch functions $\psi_n \propto
\exp(i\mu n)$ with dispersion relation $E-V_0 = 2t\cos\mu$. The
allowed energies of these extended states lie in a single band of
width $2t$, $\mid E-V_0 \mid \leq t$. In the opposite case of a
white-noise potential, where $\langle \epsilon_i \epsilon_k
\rangle=\epsilon_0^2 \delta_{ik}$ and $\langle \epsilon_n \rangle
= 0$, all the states are localized. For weak fluctuations,
$\epsilon_0^2  \ll t^2$, the Lyapunov exponent (inverse
localization length) in the Born approximation is given by
\cite{Thou}
\begin{equation}
\label{white} \gamma_0(E)= \frac{1}{l_0(E)} = \frac{\epsilon
_0^2}{8t^2 \sin^2\mu}.
\end{equation}
In this approximation, the wave function extends over many sites,
i.e., $\l_0(E) \gg 1$, and the dispersion relation remains the
same as for the regular potential, $E= 2t\cos\mu$.

\begin{figure}
\includegraphics[width = 12 cm]{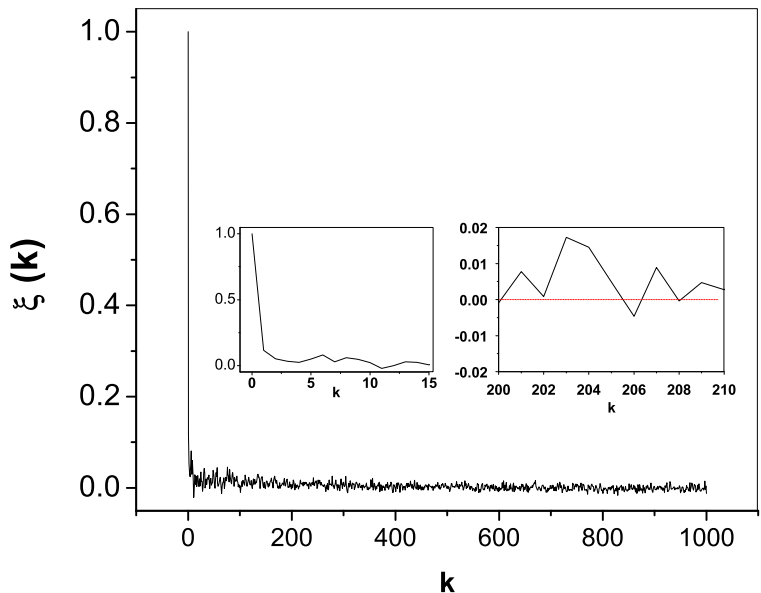}
\caption{ Binary correlation function of the sequence of
nucleotides for Human BCRA gene. The correlation function drops
from 1 at $k = 0$ to about 0.1 at $k \geq 1$, left inset.
Correlations extend to distances of a few thousands of base pairs,
decaying very slowly. An important feature of this correlation
function is close to regular oscillations about zero, right
inset.} \label{fig1}
\end{figure}

Most of the existing random potentials are neither ideally
periodic nor ideally disordered (white-noise potential). They form
a wide class of so-called {\it correlated disordered potentials}.
A generalization of Eq. (\ref{white}) for this class of potentials
was obtained in Ref. \cite{1dcorr}:
\begin{equation}
\label{loclength} \gamma(E)= \gamma_0(E)\varphi(2\mu),\,\,
\varphi(\mu )=1+2\sum\limits_{k=1}^\infty \xi (k)\,\cos \,(\mu
\,k).
\end{equation}
Here $\xi(k) = \langle\epsilon_n \epsilon_{n+k}\rangle /
\epsilon_0^2 $ is the normalized binary correlator of the
potential.
Because of the correlations the energy spectrum may contain
localized as well as extended states. In a first approximation
over $\epsilon_0^2$ the extended states occupy the intervals where
the function $\varphi(\mu)$ in Eq. (\ref{loclength}) vanishes. The
regions of localized and extended states are separated by a ``
mobility edge." For example, a sharp vertical mobility edge at
$\mu = \pi/3$ ($E=t$) appears if the correlation function decays
slowly and oscillates: $\xi (k)= (3/2 \pi k)\sin (2\pi k /3)$. In
Refs. \cite{exp1} this type of sharp mobility edge was observed in
the transmission  and reflection spectra of a microwave waveguide
with specially designed correlated scatterers. Power-law decay and
oscillations of $\xi(k)$ are the necessary (although not
sufficient) attributes of a sharp mobility edge in the energy
spectrum. We studied the correlation function of many different
DNA sequences and all of them exhibit slow decay and oscillations.
A typical correlation function is shown in Fig. \ref{fig1}. A
different approach to the Anderson transition in 1D potentials
with long-range correlations has been developed by Moura and Lyra
\cite{ML98}. It is based on the method of generation of a random
correlated sequence that is adopted from the theory of fractional
Brownian motion. The presence or absence of a mobility edge is
determined by the scaling properties of power spectrum of binary
correlation function \cite{SNN04}, but not by the binary
correlator itself.

Although Eq. (\ref{loclength}) correctly accounts for the
correlations in a {\it single-channel} random potential, it is not
appropriate for  the analysis of electron localization in real
DNA. A DNA molecule is a {\it two stranded} sequence of
nucleotides, i.e., there are two conducting channels. It is known
that the localization length strongly depends on the number of
channels in disordered chains \cite{Hein,Rom1}. Our case is even
more specific since two strands, being random in the longitudinal
direction, exhibit regular A-T and C-G matching in the transverse
direction. This key-to-lock matching between the strands strongly
affects electron transport in DNA.

To date, there have been a number of studies of the localization
length in DNA molecules. In Ref. \cite{Stanly}  an attempt was
done to obtain numerically a localization-delocalization
transition in a single-stranded binary artificial DNA sequence
with a special kind of slowly decaying correlations. However,
since in the thermodynamic limit the proposed sequence turned out
to be a regular one, the problem of extended states remains open.
In the numerical study \cite{Rom} an unexpected tendency to
delocalization with an {\it increase} of non-perturbative disorder
in an {\it uncorrelated} single-stranded DNA was observed.
Recently it was claimed that the transverse key-to-lock base
pairing by itself gives rise to a band of extended states even if
the longitudinal correlations are ignored \cite{Shulz}. This
numerical result has since been criticized using analytical
argumentation \cite{Adame}. Thus, it is now clear that for a
correct evaluation of the localization length in DNA one has to
({\it i}) use the two-stranded model; ({\it ii}) avoid
simplification of the 4-letters DNA alphabet to a binary sequence
and; ({\it iii}) account for the longitudinal correlations in both
strands as well as transversal base pairing.

In our study we use a two-channel model where the Schrodinger
equation reads \cite{Hein}
\begin{equation}
\begin{array}{c}
\label{Schr1}
    t (\psi_{1,n+1}+  \psi_{1,n-1})  + h \psi_{2,n}=
    (E-\varepsilon_{1,n}) \psi_{1,n}\,,\\

    t (\psi_{2,n+1}+  \psi_{2,n-1}) + h \psi_{1,n}=
    (E-\varepsilon_{2,n}) \psi_{2,n}\,.\\
  \end{array}
\end{equation}
Here $\psi_{1,n}$ and $\psi_{2,n}$ are the on-site wave functions
in the first and second chain, respectively, and
$\varepsilon_{1,n}$ and $\varepsilon_{2,n}$ are the on-site
potentials. The hopping parameters $t$ and $h$ determine the
inter- and intra-strand coupling.

In the case of a periodic potential there are two momenta $\mu_1$
and $\mu_2$ for each energy $E$. They are given by two dispersion
relations, $E=2t \cos\mu_{1,2} \pm h$. Here we consider the case
of a band structure when the two propagating channels overlap.
This happens if $ h <2t$. The band of allowed energies spreads
from $-2t-h$ to $2t+h$. Both channels are propagating (i.e.
$\mu_1$ and $\mu_2$ are real) for the energies $ \mid 2t - h \mid
< E $.

\section{Localization length}
For calculation of the localization length we use the perturbation
theory approach developed for two- and three-channel waveguide in
Ref. \cite{Hein}. The localization length is defined by the
following formula
\begin{equation}
\label{conductance} \gamma(E)=
l^{-1}(E)=-\lim_{N\rightarrow\infty} \frac{1}{2N} \langle\ln
Tr{(\hat t \hat t^{\dag})}\rangle,
\end{equation}
where $\langle\ldots\rangle$ denotes averaging over disorder and
$\hat t$ is $2 \times 2$ transmission matrix. The transmission
matrix $\hat{t}$ that enters into the Landauer formula $g =
(2e^2/h)Tr(\hat{t}\hat{t}^{\dagger})$ is calculated as a product
of $N$ on-site transfer-matrices. The transmission matrix is
calculated in the linear (Born) approximation over weak disorder
$\langle\varepsilon_1^2\rangle, \langle\varepsilon_1^2\rangle \ll
t^2$. The results reported here are based on the following formula
for the Lyapunov exponent \cite{Bag}:
\begin{eqnarray}
\label{loclength2}
 \gamma(E) = \frac{\varepsilon_{1}^2}{64t^2}
\left[\frac{\varphi_{11}(2\mu_1)}{\sin^2 \mu_{1}} +
\frac{\varphi_{11}(2\mu_2)}{\sin^2 \mu_{2}}+
\frac{2\varphi_{11}(\mu_1 + \mu_2)}{\sin\mu_{1}\sin\mu_{2}}
\right] \nonumber \\
+ \frac{\varepsilon_{2}^2}{64t^2}
\left[\frac{\varphi_{22}(2\mu_2)}{\sin^2 \mu_{2}} +
\frac{\varphi_{22}(2\mu_1)}{\sin^2 \mu_{1}}+
\frac{2\varphi_{22}(\mu_1 + \mu_2)}{\sin\mu_{1}\sin\mu_{2}}
\right] \nonumber \\
+ \frac{\varepsilon_{12}}{32t^2}
\left[\frac{\varphi_{12}(2\mu_1)}{\sin^2 \mu_{1}} +
\frac{\varphi_{12}(2\mu_2)}{\sin^2 \mu_{2}}-
\frac{2\varphi_{12}(\mu_1 + \mu_2)}{\sin\mu_{1}\sin\mu_{2}}
\right] .
\end{eqnarray}
In the two-channel model electron localization occurs due to
backscattering processes in both channels with intra-channel
momenta transfers $2\mu_1$ and $2\mu_2$. There is also
inter-channel scattering with momentum transfer $\mu_1 + \mu_2$.
Accordingly, there are terms $\varphi_{11}(2\mu_1)$,
$\varphi_{22}(2\mu_2)$, and $\varphi_{12}(\mu_1+\mu_2)$ in Eq.
(\ref{loclength2}). The functions $\varphi_{ij}$ are expressed
through three binary correlators $\xi_{ij}$, similarly to Eq.
(\ref{loclength}):
\begin{equation}
\label{phi} \varphi_{ij}(\mu)=1+2\sum_{k=1}^{\infty}\xi_{ij}(k)
\cos(\mu k)\,\,\,\,\,i,j=1,2.
\end{equation}
These functions, $\xi_{11}$, $\xi_{22}$ and $\xi_{12}$,
characterize the intra- and inter-channel correlations
respectively:
\begin{equation}
 \begin{array}{cc}
 \langle\varepsilon_{1,n} \varepsilon_{1,n+k}\rangle =  \varepsilon_{1}^2
 \xi_{11}(k),\,\,
  \langle\varepsilon_{2,n}
 \varepsilon_{2,n+k}\rangle = \varepsilon_{2}^2 \xi_{22}(k), \\
 \langle\varepsilon_{1,n} \varepsilon_{2,n+k}\rangle = \varepsilon_{12} \xi_{12}(k).
 \end{array}
 \end{equation}
Here the mean value $ \varepsilon_{12} = \langle\varepsilon_{1,n}
\varepsilon_{2,n}\rangle $ can be either positive or negative,
unlike always positive variances $\varepsilon_{1,2}^2$. Equation
(\ref{loclength2}) is valid if both channels are propagating,
i.e., the wave numbers $\mu_1$ and $\mu_2$ are real. If one of the
channels becomes evanescent it is replaced by the eq. (33) from
Ref.  \cite{Bag}.
\begin{equation}
\label{loclength3} \gamma(E)=\frac{1}{32 \sin^{2}
\mu_1}\left[\varepsilon_{1}^{2}\varphi_{11}(2\mu_1) +
\varepsilon_{2}^{2}\varphi_{22}(2\mu_1) + 2
\varepsilon_{12}\varphi_{12}(2\mu_1)  \right].
\end{equation}
At the transition points when $E=E_c=\mid 2t-h \mid$, one of the
denominators in Eq. (\ref{loclength2}) vanishes ($\sin
\mu_{1,2}=0$) and the Born approximation fails.

We apply Eq. (\ref{loclength2}) to a two-stranded DNA molecule.
Among a huge number of chemical and physical characteristics of a
DNA molecule, we need here only the ionization potentials for each
nucleotide, $\epsilon_A$, $\epsilon_T$, $\epsilon_C$, and
$\epsilon_G$, and two hopping amplitudes, $t$ and $h$. Unlike
previous studies, we develop here an analytical two-channel
approach, which accounts for intra- and inter-channel
correlations. Therefore, we do not simplify a two-stranded DNA
sequence to a binary sequence, using a coarse-graining procedure.
From this point of view, our approach is much more close to
reality.

The length of a DNA sequence may reach $~10^6-10^9$ base pairs. In
such a long disordered chain all the states are localized and a
DNA molecule does not conduct. Much shorter segments may, however,
exhibit very different behavior \cite{muta,Enders}. This means
that the conducting properties of a DNA molecule vary along the
sequence of nucleotides and explains a wide spectrum of conducting
properties obtained in experiments; see in the references in
\cite{Enders}. The physical characteristics, like the ionization
potential and hopping amplitudes, are independent of the position
of a given nucleotide in the sequence. The only characteristics
which may change along the sequence are the correlation functions.
The exons and introns store different kind of information and this
affects the correlators. Thus, the localization length and the
conductance of a given segment of a DNA molecule are directly
related to the genetic information stored in this segment.
Equation (\ref{loclength2}) is a mathematical manifestation of
this fact.

\section{Numerical results for localization length}

In order to demonstrate the inhomogeneities in the conductivity of
the DNA molecules we studied the localization length along the
exons and introns. Exons are the parts where the genetic
information is written and introns are the parts without apparent
information for protein synthesis. The introns occupy a larger
part of the DNA sequence of higher eukaryotes than do the exons.
For procaryotes the situation is the opposite. From the point of
view of ``quality" of the carried information the introns and
exons are the most different segments and one may expect very
distinct localization properties to exist in these segments. It
was recently shown that the melting of exons and introns also
occurs in a different way \cite{therm}.

\begin{figure}
\includegraphics[width = 12 cm]{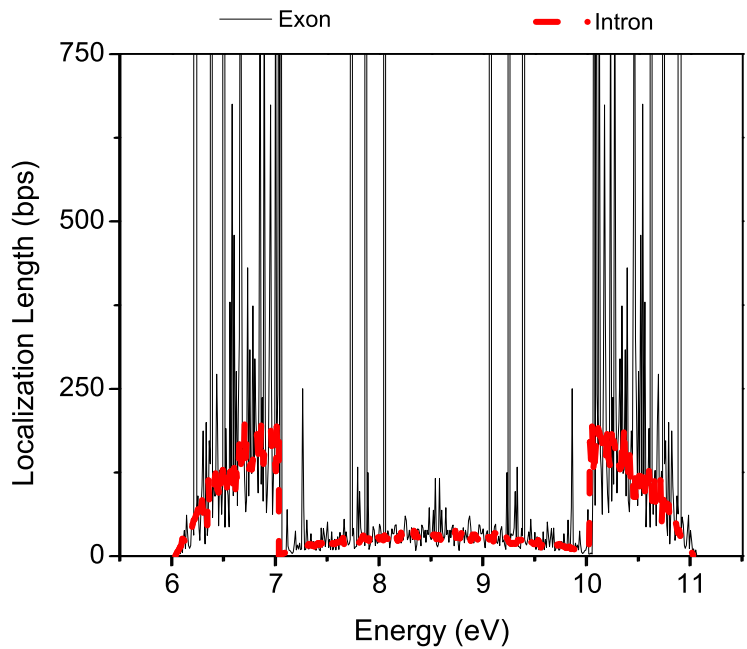}
\caption{ Color online. Localization length vs energy for the
Human BRCA gene measured in the number of base pairs. The length
of the exon (intron) is 2120 (10421) base pairs(bps). The results
for exon and intron are shown by black and grey (red) lines
respectively. The values of the hopping parameters are $h=0.5$ eV
and $t=1$ eV. The two channels are propagating if $6.6 < E < 10.4$
eV. One of the channels becomes evanescent in two symmetric
regions, $10.4< E < 11.4$ eV and $5.6<E<6.6$ eV, of the width of
$2h=1$ eV.} \label{LLBRCA}
\end{figure}

\begin{figure}
\includegraphics[width = 12 cm]{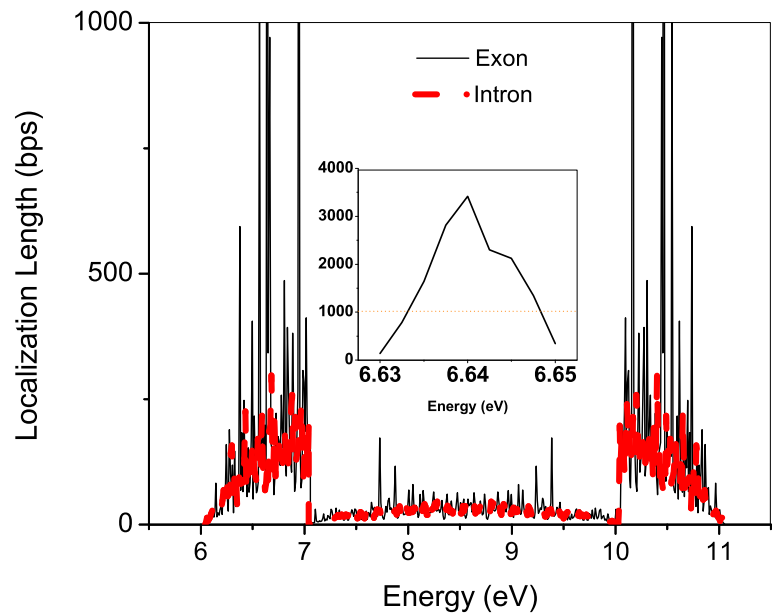}
\caption{ Color online. Localization length for the Human ADAM10
gene. The length of the exon (intron) is 1030 (31752) base
pairs(bps). Inset shows the fine structure of one of the peaks.}
\label{LLADAM10}
\end{figure}

\begin{figure}
\includegraphics[width = 12 cm]{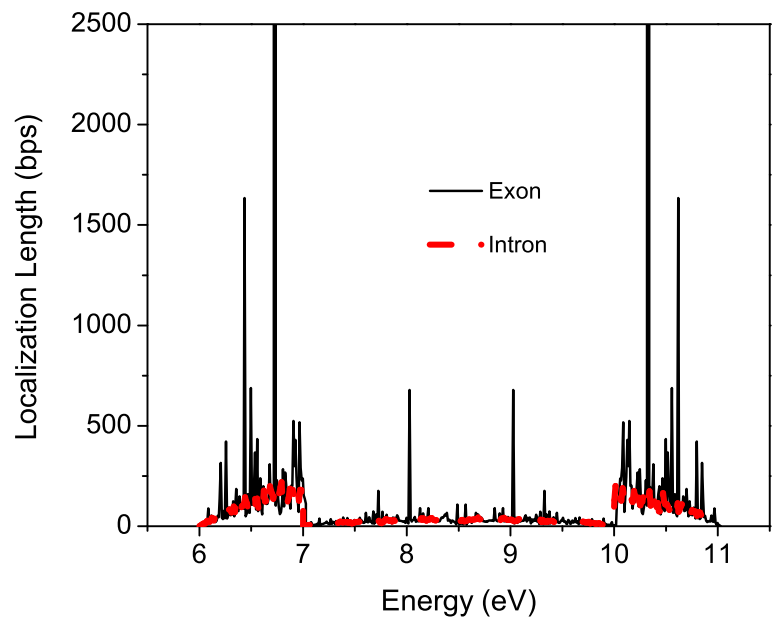}
\caption{ Color online. Localization length vs energy for the
Human SNAP29 gene. The length of the exon (intron) is 2141 (21701)
base pairs(bps).} \label{LLSNAP29}
\end{figure}

\begin{figure}
\includegraphics[width = 12 cm]{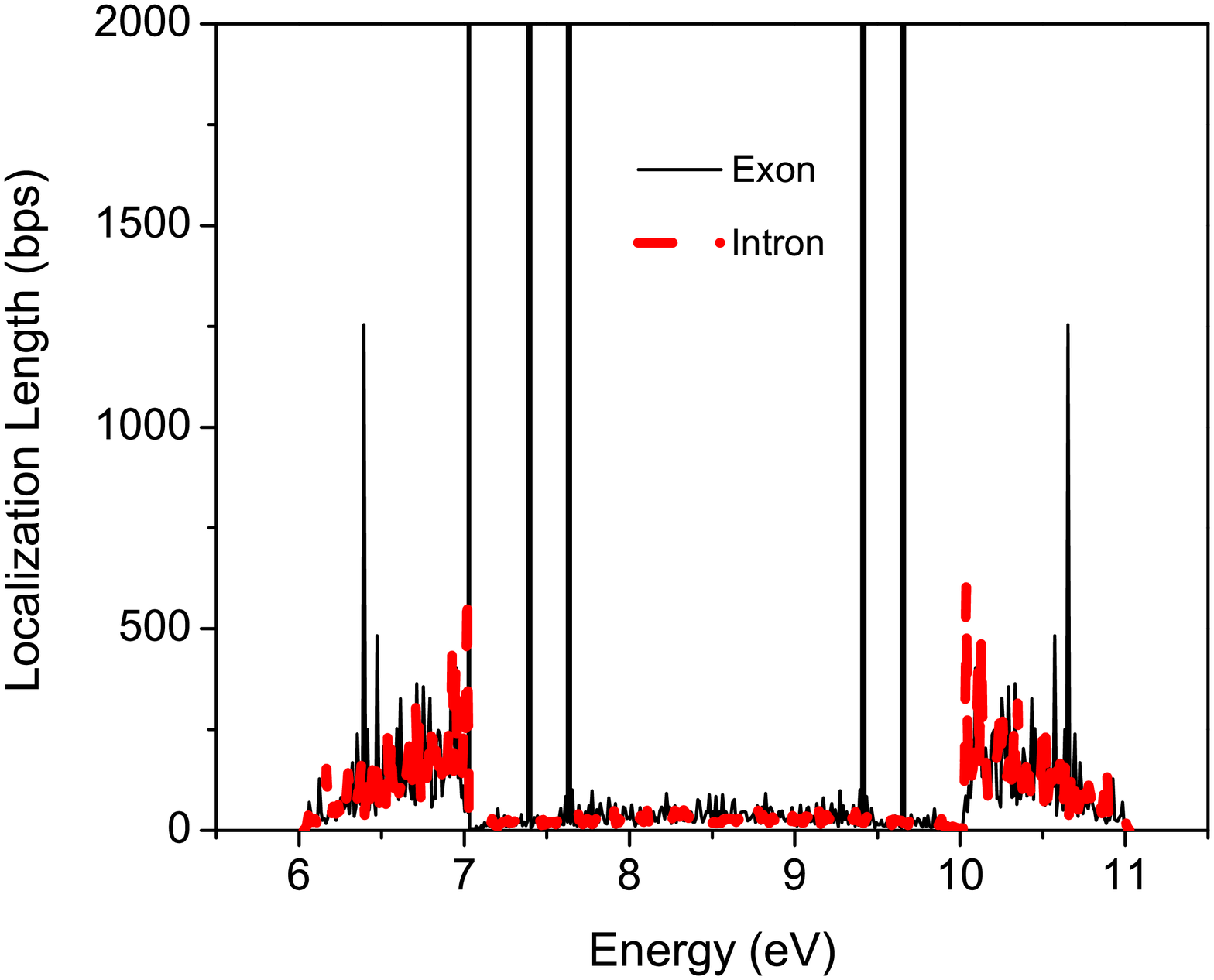}
\caption{ Color online. Localization length vs energy for the
Human SUHW1 gene. The length of the exon (intron) is 1963 ( 4405)
base pairs(bps).} \label{LLSUHW1}
\end{figure}

We use Eqs. (\ref{loclength2}) and (\ref{loclength3}) for
numerical calculation of the localization length. Here we give the
results for the following human DNA molecules: BRCA, ADAM10,
SNAP29, and SUHW1. The results are shown in Figs. \ref{LLBRCA} -
\ref{LLSUHW1} where we plot the localization length vs. electron
energy for the exon and intron segments. The parameters of
nucleotide site energies and the hopping amplitudes are the same
for all these figures. Thus, the very different patterns shown in
the figures represent different information codes in different
DNA's.

For most of the energies the localization length inside the exon
region exceeds by order of magnitude the localization length
inside the intron region. This confirms, by implication, the fact
that very different kinds of information are coded in these
regions. The vertical axis for each figure is cut off
approximately at the length of the corresponding exon region.
There are many peaks in the exon regions with the hight that
exceeds much the vertical scale, i.e. the states within these
peaks are extended. Unlike this, in the intron regions all the
states are well-localized. The density of the peaks in Figs.
\ref{LLBRCA} and \ref{LLADAM10} is much higher than that in Figs.
\ref{LLSNAP29} and \ref{LLSUHW1}. Most of the peaks are situated
in the region of energy where one of the channels is evanescent.
Similar sharp peaks in the transmission of the exon regions of Y3
DNA have been numerically obtained in Ref. \cite{Shih} for a
single stranded DNA. It turns out that this feature is very robust
since in that study a single-stranded model of DNA was used.

The fine structure of one of the peaks is shown in the inset of
Fig. {\ref{LLADAM10}}. Since the peaks are of a finite width
($\sim 20$ meV), they are narrow bands of extended states, but not
the discrete resonant states predicted and observed in random
dimers \cite{dimer}. The nature of resonant tunnelling in random
dimers is due to short-range correlations in contrast with
specific long-range correlations which are necessary for existence
of a continuous band of extended states. In the case of a single
channel the width of the band of extended states can be controlled
by the parameters of the binary correlator $\xi(k)$ in Eq.
(\ref{loclength}). In particular, wide and narrow bands of the
extended states have been observed in the experiments with
single-mode microwave waveguides \cite{exp1}. For the two-channel
system the relation between the positions of the mobility edges
and the explicit form of the binary correlator is not known.  One
may expect that such relation is determined by the relative phase
shifts between the Fourier components of the oscillatory
correlators $\xi_{ij}(k)$. It is worth mentioning that short- and
long-range correlations lead not only to different localization
properties but also to very different classical as well as quantum
diffusion in DNA \cite{Lyra}.

\begin{figure}
\includegraphics[width = 8 cm]{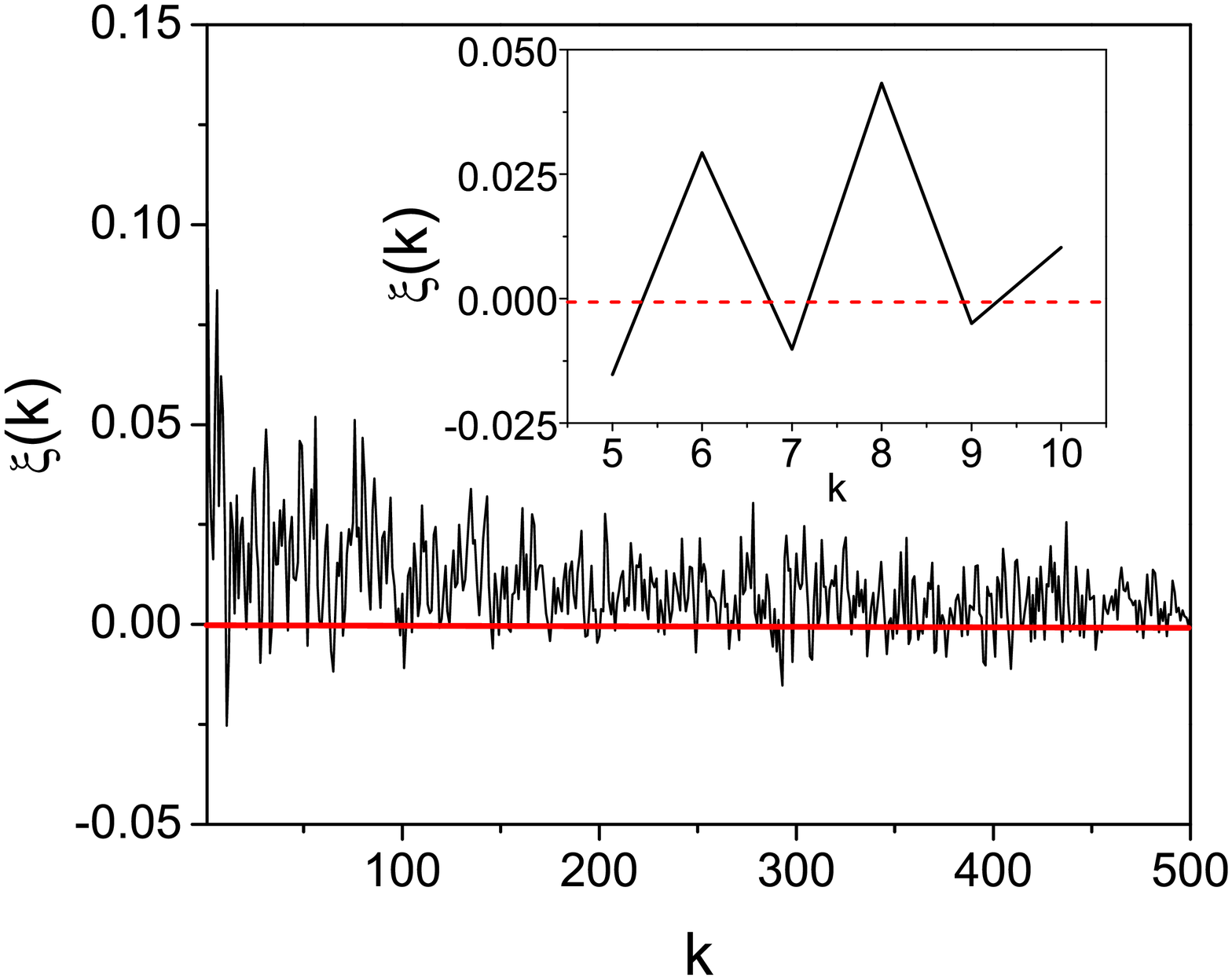}
\includegraphics[width = 8 cm]{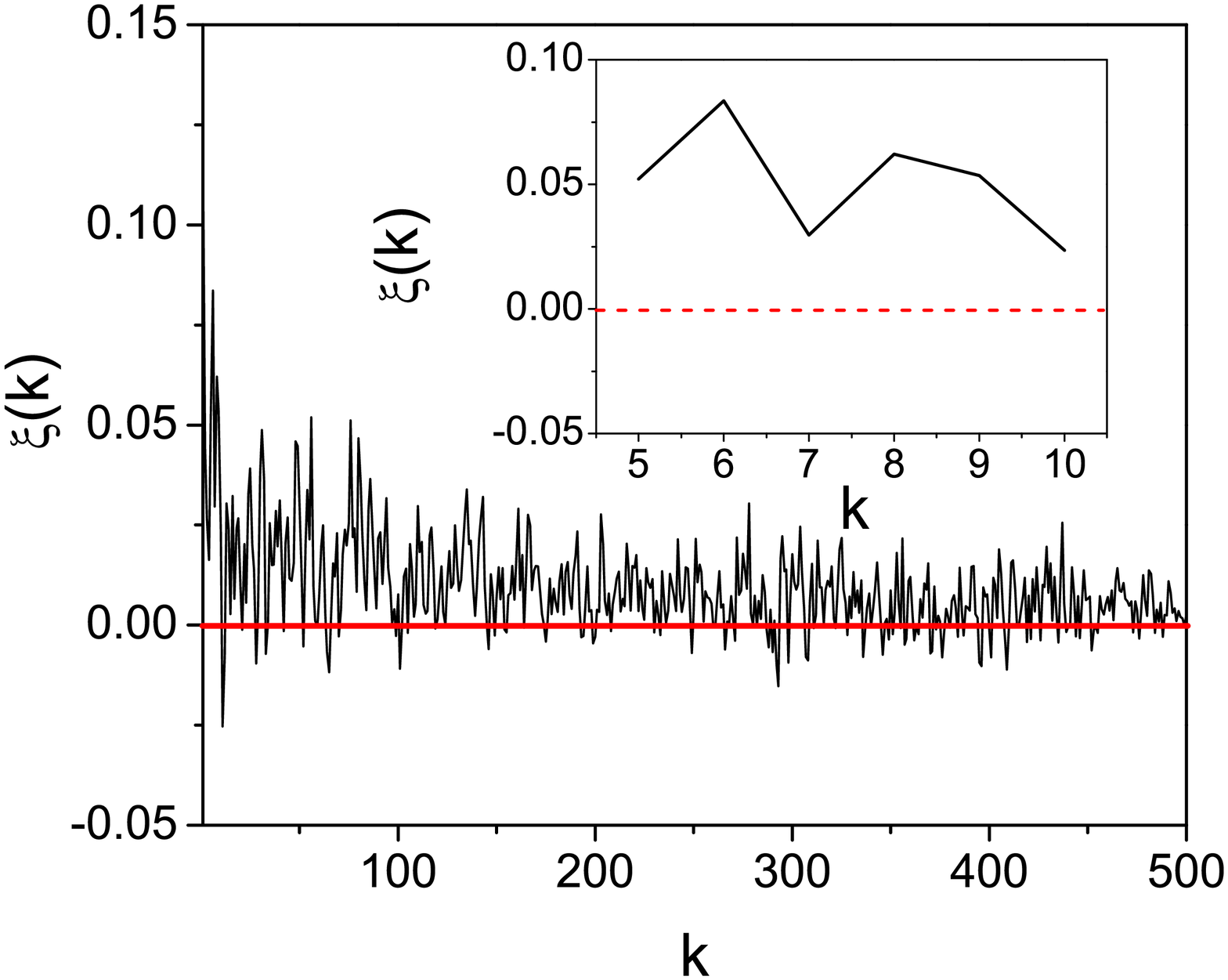}
\caption{ Color online. Binary correlator for the exon (left
panel) and intron (right panel) regions of the Human SNAP29 gene.
Inserts show local behavior of the correlators within small
intervals of $k$.} \label{CorrSNAP29}
\end{figure}

A pattern $l(E)$ is a particular fingerprint of a given DNA
sequence and it can be used, in principle, for classification of
DNA molecules. In the previous studies (see, e.g.,
\cite{Peng,Usat}) the DNA sequences have been characterized by
scaling exponent of the corresponding random walk. We consider
that the inverse localization length (\ref{loclength2}) is more
convenient since it characterizes a well defined physical property
-- electrical resistivity. Moreover, Eq. (\ref{loclength2})
establishes a qualitative relation between the localization length
and the informational characteristic (binary correlators) of the
DNA sequence. At the same time the binary correlators by
themselves are not very illustrative. In particular, the plots of
the correlators for exon and intron regions look very similar, see
Fig. \ref{CorrSNAP29}, although these plots, of course, contain
the same information about the DNA sequence as the plots for the
Lyapunov exponents do. It is clear that the presence or absence of
the bands of the extended states is determined by subtle
interference among the Fourier harmonics of the functions
$\varphi_{ij}(\mu)$ given by Eq. (\ref{phi}).

The Lyapunov exponent (\ref{loclength2}) depends on $\epsilon_A$,
$\epsilon_T$, $\epsilon_C$, and $\epsilon_G$ as well as on the
hopping amplitudes $t$ and $h$. Since the values of $t$ and $h$
are not well established experimentally, we repeated the
calculations for different values of the hopping amplitudes,
$0.1<h<0.5$ and $0.7<t<1$. Since our analytical approach is valid
only in the region where the perturbation parameters $\epsilon_1/
t$ and  $\epsilon_2/ t$ are small, we cannot extrapolate our
results to the region where $t<0.5$. The patterns for the Laypunov
exponents do not change essentially with variation of the
parameters $t$ and $h$. The delocalized states do not disappear
but the positions of the mobility edges are slightly displaced.


\section{Conclusions} In our study of the double-stranded model of
DNA we observed much longer localization length in exon than in
intron regions for practically all the allowed energies and for
all randomly selected DNA sequences. Through statistical
correlations of the nucleotide sequence making up a DNA molecule,
we relate this persistent difference to qualitatively different
information stored by exons and introns.

For each DNA the pattern $l^{-1}(E)$ is unique fingerprint  and
can be used for identification of DNA's. All presented results
confirm the suggestion that the localization length in DNA is
determined by specific long-range correlations between the
nucleotides  and not by a particular choice of control parameters
of the model.

The conducting properties of DNA have attracted much attention
since DNA may be used in electronic devices \cite{appl,appl1}. We
hope that our approach and results can be very useful for further
theoretical and experimental studies of the electrical and optical
properties of DNA.

\section{Acknowledgement} This work is supported by the US
Department of Energy grant \# DE-FG02-06ER46312.

\end{document}